\documentclass[prf,amsmath,amssymb,aps]{revtex4-2}
\usepackage{amsfonts}
\usepackage{amsmath}
\usepackage{amsthm}
\usepackage{graphics}
\usepackage{graphicx}
\usepackage{subfigure}
\usepackage{amssymb}
\usepackage{graphics}
\usepackage{graphicx}
\usepackage{epstopdf}
\usepackage{color}
\usepackage{subfigure}
\usepackage{pgfplots,tikz}
\usepackage{url}
\usepackage{soul}
\usepackage{enumitem}
\usepackage{lipsum}
\usepackage{hyperref}
\usepackage{cleveref}
\usepackage{comment}

\def\x{{\bf x}}

\def\k{\textbf{k}}

\begin{document}

\title{Initial evolution of 3D turbulence en route to the Kolmogorov state: emergence and transformations of coherent structures, self-similarity and instabilities}

\author{Giorgio Krstulovic}
\affiliation{Universit\'{e} C\^{o}te d'Azur, Observatoire de la C\^{o}te d'Azur, CNRS, Laboratoire Lagrange, Boulevard de l'Observatoire CS 34229 -- F 06304 Nice Cedex 4, France}
\author{Sergey Nazarenko}
\affiliation{Universit\'{e} C\^{o}te d'Azur, CNRS, Institut de Physique de Nice - INPHYNI, Parc Valrose, 06108 Nice, France}

\begin{abstract}
    In this work, we study numerically the temporal evolution of an initially random large-scale velocity field under governed by the hyperviscous incompressible Navier-Stoke equations. Three stages are clearly observed during the evolution. First, the initial condition development is characterized by a spectrum evolving in a self-similar way with a wave number front $k_*(t)$ propagating toward high values exponentially in time. This evolution corresponds to the formation and shrinking of think vortex pancakes exponentially in time, as it has been previously reported in simulations of the incompressible Euler equations. At the second stage, pancakes become unstable, rolling up on the edges and breaking up in the middle, leading to the emergence of \emph{vortex ribs}-- quasi-periodic arrangements of vortex filaments. Those filaments then twist creating structures akin to ropes. At the last stage, a fully developed turbulent state is observed, characterized by a Kolmogorov energy spectrum and exhibiting a decay law compatible with standard turbulent predictions in the case of an integral scale saturated at the scale of the box.
\end{abstract}

\maketitle

\section{Introduction}
Traditionally, studies of hydrodynamic turbulence focus on statistically steady states. These include testing the degree of validity of the Kolmogorov theory and investigating deviations from this theory \cite{frisch1995turbulence}. This is related to the view that turbulence properties are most universal, and therefore most interesting, when a stationary state is reached. However, non-stationary turbulence is also interesting and important, and it deserves greater attention. Particularly interesting is the temporal evolution en route  to the steady state. Here, some unexpected nontrivial behavior was discovered for phenomenological descriptions of turbulence: the Leith and the eddy-damped quasi-normal Markovian (EDQNM) models.
The Leith model is a nonlinear singular diffusion equation for the energy spectrum \cite{leith1967diffusion,connaughton2004warm}, whereas the EDQNM model is a nonlinear integro-differential equation \cite{orszag_1970,GNMCh2014}. It was found, first for the Leith model and later for the EDQNM model, that turbulence initially excited in a finite range of scales develops a power-law spectrum  with an exponent $x_*$ which is different from Kolmogorov's $-5/3$ behind the front propagating toward higher wave numbers. Namely, for the Leith model $x_*$ was found to be $\approx -1.85$ \cite{connaughton2004warm,GNMCh2014} and for the EDQNM closure $\approx-2$ \cite{BOS}.
It is interesting that such an exponent is model-dependent but independent of the shape of the initial spectrum.  The Kolmogorov spectrum in both of these models forms after time $t_*$ when the initial spectral front reaches the dissipative scale: it propagates as a wave in the $k$-space propagating from the dissipative scale toward the initial scale, i.e. from large to small wave numbers. This evolution scenario is typical for all finite-capacity turbulent systems, i.e. systems whose constant-flux (Kolmogorov-type) spectra correspond to a finite physical-space density of the cascading invariant (e.g. energy) when the dissipation coefficient (e.g. viscosity) tends to zero. Besides the Leith model and EDQNM, this behavior was also observed for the Magneto-Hydrodynamic (MHD) turbulence \cite{galtier2000} and turbulence described by the Burgers equation \cite{SULEM}, and later for a great variety of Wave Turbulence systems
\cite{Thalabard2015}.

Analytical treatment is naturally easier for the differential (e.g. Leith) models than for the integro-differential ones (e.g. EDQNM). In the Leith model case, the pre-$t_*$ and the post-$t_*$ stages to the self-similar solutions of the second and the third kinds respectively. In the  self-similarity of the second kind the similarity coefficients are found by solving a nonlinear eigenvalue problem, whereas in the third kind self-similarity they are fixed by the previous self-similar stage. It  was shown that the anomalous power-law exponent  c for the   pre-$t_*$ spectrum is related is a heteroclinic bifurcation of a dynamical system associated with the self-similar class of solutions. The  post-$t_*$ stage (with the Kolmogorov scaling invading the $k$-range from right to left) has the similarity coefficients inherited from  the pre-$t_*$. Similar behavior is typical for the integro-differential models; for example it was demonstrated for the kinetic equation of the weak wave MHD turbulence  \cite{galtier2000}.

A revealing picture arises when analyzing a similar scenario in the Burgers equation model. Since the Burgers equation is integrable, its analysis can be done fully rigorously. It was found \cite{SULEM} that the  pre-$t_*$ scaling is associated with formation of the so-called pre-shock: a cubic-root singularity arising in the physical at the wave breaking moment.  The post-$t_*$ scaling is related to the shock singularity giving the Burgers power-law spectrum which is an analog of the Kolmogorov spectrum.
It is tempting to extend such a two-singularities scenario to the hydrodynamic turbulence. 
However, the closure models which only involve the spectrum do not allow to study singularities because they do not contain information about the phases of the Fourier amplitudes of velocity. On the other hand, a two-singularity scenario was indeed discussed in \cite{Brachet92}, who used pseudo-spectral direct numerical simulations (DNS) of the Euler equations at resolution $256^3$  and observed  that pancake-like vorticity structures form at the initial (inviscid) stage of the turbulent evolution. Appearance of the vortex pancakes, and accompanying it power-law with exponent $\approx-4.5$,  turned out to be a rather robust process--it was observed in runs with initial conditions in the form of the Taylor-Green vortex as well as in runs with random initial conditions.  The pancakes were observed to be catastrophically shrinking in thickness--a process which should eventually be arrested by viscosity. They further suggested (but not demonstrated numerically) that the pancakes should become unstable at the moment when their thickness becomes comparable to the dissipative scale. Such a instability should lead to formation of another type of quasi-singular vortex structures--thin intense vortex filaments. Later in \cite{Passot1995}, pseudo-spectral $256^3$ DNS of the Navier-Stokes equations was performed using the same random-mode initial conditions as in \cite{Brachet92}. Presence of viscosity and use of a state of the art (for that time) visualization technique allowed them to observe instability of the vortex pancakes leading to their breakup into filamentary vortices. Also, the authors of 
\cite{Passot1995} have suggested that the prototype for the observed vortex pancakes are the so-called Burgers vortex  layers in which an interplay of an external strain and the viscosity results in a stationary sheet-like vortex structure
\cite{saffman1995vortex}. They performed a linear analysis of instability of the Burgers vortex  layer in a way which was previously used to demonstrate instability of the mixing layers in \cite{lin84}. Later such a linear instability analysis was further developed in \cite{Beronov1996}.

 The process of formation of shrinking vortex pancakes in evolving turbulence was recently reproduced numerically in high-resolution  DNS of the Euler equations: at resolution $486 \times 1024 \times 2048$  in \cite{agafontsev2015} and $972 \times 2048\times 4096$ -- in \cite{agafontsev2017}.  Self-similarity in evolving pancakes was observed and characterized. However, no break-up of the vortex pancakes into filamentary was observed. Later, the same group of authors claimed that no Kelvin-Helmholtz type instability is possible for the rapidly shrinking vortex pancakes. These observations are in agreement with the suggestion that including dissipation is essential for the vortex pancake instability because it is dissipation that slows down the pancake collapse and makes them behave like quasi-stationary Burgers vortex layers. It was also claimed in \cite{agafontsev2015,agafontsev2017} that the shrinking vortex pancakes give rise to appearance of the Kolmogorov -5/3 spectrum.
  
The present paper is devoted to a study of the initial evolution of hydrodynamic turbulence by DNS of the Navier-Stokes equations with hyper-viscosity using a pseudo-spectral method and random wave initial conditions similar to the ones used in \cite{Brachet92}. Our motivation is twofold: we would like to check if the Fourier-space self-similarity (previously observed in the Leith and the EDQNM closures) is also realized in DNS of hydrodynamic turbulence, as well as to examine, at high resolution, realizability of the previously conjectured two-singularity scenario in where the initial evolution first leads to formation of shrinking vorticity pancakes which subsequently become unstable and break up into thin quasi-singular vortex filaments.

We find that the initial evolution stage is characterized by a spectrum evolving in a self-similar way with a front $k_*(t)$ propagating toward high $k$'s exponentially in time. 
Note that the exponential shrinking of the minimal size $\delta(t) \sim 1/k_*(t)$ was also reported in \cite{Brachet92} and later in \cite{agafontsev2015} and  \cite{agafontsev2017}. We observe that the spectral front reaches the maximum wavenumber in a finite time $t_* \approx 1.24$. On the other hand, at the present available resolution we were not able to detect any front acceleration which could be interpreted as a change from the exponentially shrinking regime to a finite-time blowup behaviour. Thus, our tentative observation is that the value of $t_*$ is resolution dependent and would increase if the inertial range of wavenumbers was widened, and that no evidence for the finite time singularity formation could be seen.
  
In the physical space, like  previously reported in \cite{Brachet92,agafontsev2015} and  \cite{agafontsev2017}, at the early stage we  see formation of  vorticity pancakes whose thickness is shrinking. After the minimal thickness have reached the dissipative scale, the pancakes go unstable: they roll on the edges  and break up in the middle parts into "vortex ribs"--quasi-periodic sets of intense vortex filaments. Thus, the scenario of two consecutive (quasi-)singularities is indeed observed. Note that the pancakes form at before $t_*$ and persist/grow well beyond $t_*$. Their instability and breakup into quasi-periodic arrays of co-directed vortex ribs starts much later, at $t \gtrsim 3.5 t_*$. After forming, the vortex ribs pull together and twist forming vortex ``ropes".
The Kolmogorov spectrum does indeed form at the later stage, $t \gtrsim 4.6 t_*$, after the vortex ribs and ropes become numerous and entangled, forming a rather chaotic vorticity field. Note that, in contrast to what was claimed in \cite{agafontsev2015,agafontsev2017} the Kolmogorov spectrum formation 
occurs much later than the shrinking pancaked form and disappear. In fact, no spectrum numerical result for the energy spectrum was presented in \cite{agafontsev2015,agafontsev2017} to support their claim about the Kolmogorov spectrum forming at the shrinking pancake stage.

We believe that the mechanism of the pancake instability in the system with hyperviscosity is similar to the Burgers layer instability in the system with regular kinematic viscosity previously considered in \cite{Passot1995}.
Respectively, we believe that the Navier-Stokes system with 
normal viscosity should behave in a similar way to what we observe in the Navier-Stokes system with hyperviscosity considered in our paper. Namely, the route to fully developed turbulence could be roughly divided into three stages: (i) vortex pancake formation and exponential shrinking (ii) pancake roll-ups at the edges and instability in the middle parts leading to the vortex ribs, and (iii) twisting of vortex ribs and forming vortex ropes with further entanglement of the rib and rope structures and formation of a fully developed chaos in the vorticity field. This latter stage is characterised by formation of a Kolmogorov scaling in the turbulent energy spectrum.

\section{Mathematical model and numerical setup}

We consider $3$D decaying solutions ${\bf u}(\x,t)$ of the incompressible hyper-viscous Navier-Stokes equations:
\begin{eqnarray}
\frac{\partial {\bf u}}{\partial t}+{\bf u} \cdot\nabla {\bf u}&=&\nabla p-\nu (-\Delta)^\alpha {\bf u}\\
\nabla\cdot{\bf u}&=&0.
\end{eqnarray}
The total energy and total dissipation of the system are given by 
\begin{equation}
    E(t)= \frac{1}{2V}\int|{\bf u}|^2\mathrm{d}{\x}, \quad {\rm and}\quad \epsilon(t)=\frac{\nu}{V}\int {\bf u} (-\Delta)^\alpha {\bf u}\mathrm{d}{\x}=\nu \sum_{\k}  |\k|^{2\alpha}|\widehat{ {\bf u}}(\k)|^2,
\end{equation}
where $\widehat{ {\bf u}}(\k)$ is the Fourier transform of ${\bf u}$. They naturally obey the balance equation $\frac{d E}{dt}=-\epsilon(t)$.

We consider the evolution of a random large-scale initial condition, defined as
\begin{equation}
\widehat{ u}_i(\k,0)= \sum_{|\k|<k_{\rm I}} \frac{\sqrt{C}}{k}e^{i \phi_k}, \hspace{1cm} i=1,2,3.
\end{equation}
The phases $\phi_k$ are chosen randomly and $C$ is a normalization constant such that $E_0=E(t=0)$.

In the numerical simulations presented in this work we have set $\alpha=4$, $k_{\rm I}=1.75$ and $E_0=1$. The cubic domain is of size $2\pi$. We adapted the standard pseudo-spectral code GHOST [\url{https://github.com/pmininni/GHOST}]] and used resolutions of $1024^3$ and $2048^3$ colocation points with the hyper-viscous dissipation coefficient set to $\nu=4\times 10^{-19}$ and $\nu=1.52\times 10^{-21}$, respectively. Both simulations are started from exactly the same initial conditions (same large-scale Fourier modes). Run at resolution $2048^3$ is evolved up to $t=2$ and it used to perform a fine analysis of the energy spectrum at early stages. Run at resolution $1024^3$ goes up to $t=10$ and allows us to perform visualization of the enstrophy field and to study later stages of the evolution. 

\begin{figure}[h!]
    \includegraphics[width=.99\textwidth]{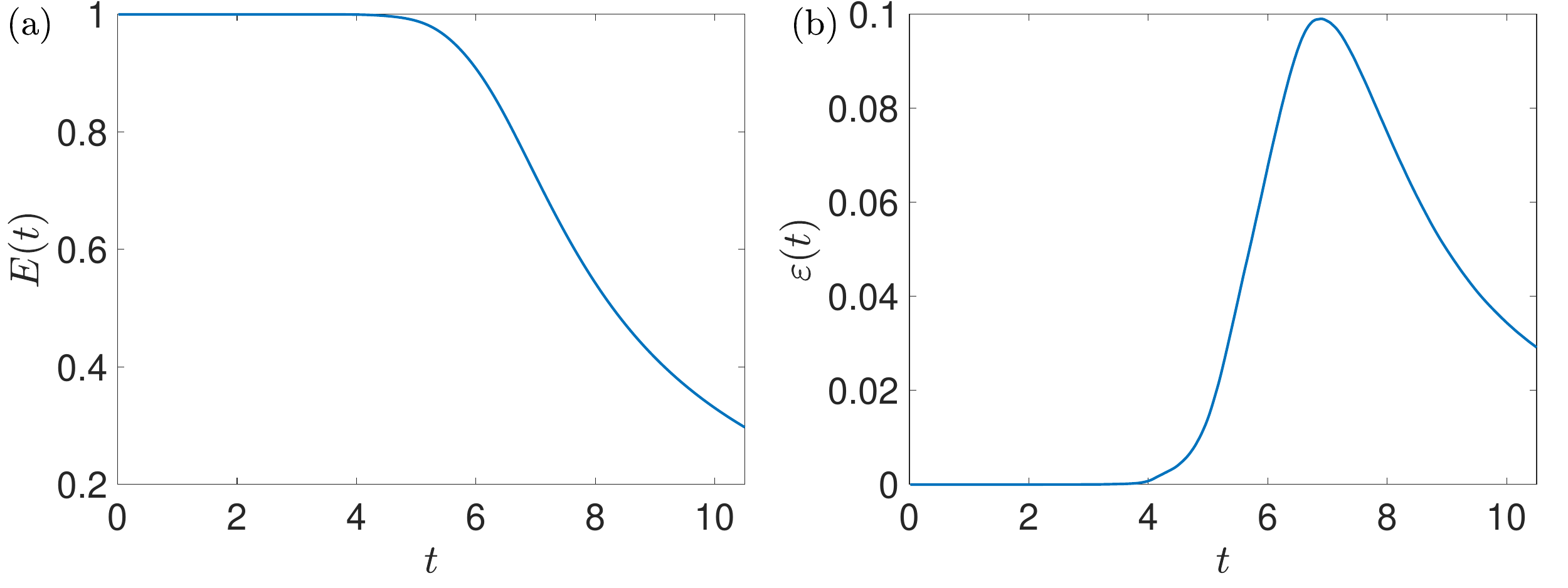}
    \caption{Temporal evolution of the energy and energy dissipation. Run at resolution $1024^3$.}
    \label{Fig:EnerAndDisp}
\end{figure}

The temporal evolution of energy and dissipation are shown in Fig. \ref{Fig:EnerAndDisp}.
We observe three stages which will be reported in the following. In early evolution (stydied in section \ref{sec:early}) dissipation plays no role, and the dynamics is effectively driven by the Euler equation ($\nu=0$). As we will se later, this phase ends when the propagating spectrum reach the viscous scale at $t^*\approx 1.225$. During the second stage, between $t^*$ and the maximum of dissipation around $t=7$ dissipation comes in, and different structures appear in the flow (see section \ref{sec:int}). Finally, closely after the maximum of dissipation turbulence is fully developed exhibiting Kolmogorov scaling (see section \ref{sec:last}).

\section{Early stages: inviscid evolution before $t_*$}
\label{sec:early}

\begin{figure}[h!]
    \includegraphics[width=.99\textwidth]{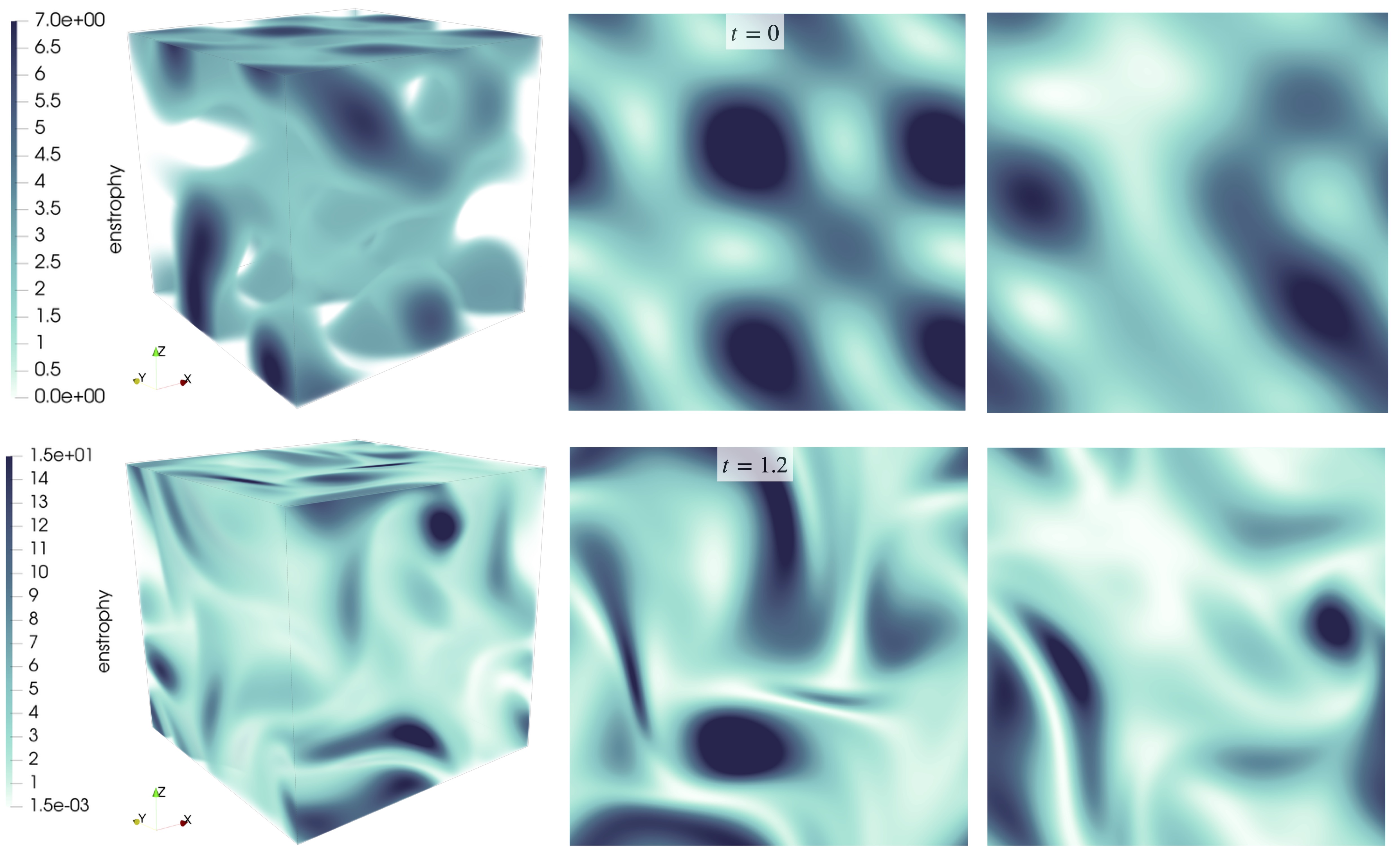}
    \caption{Visualizations of the enstrophy field at $t=0$ (top row) and $t=1.2$ (bottom row). The left panels show 3D visualizations of the field, whereas the center and right columns displays cuts of the $xy$- and $xz$-planes. Run at resolution $1024^3$.}
    \label{Fig:Visu_early}
\end{figure}

The initial condition consists on large scale structures, oriented randomly in the box. Visualizations of the initial condition are displayed in Fig.\ref{Fig:Visu_early} (top row).
During the evolution, vorticity blobs shrink, creating thinner and thinner vortex sheets, following a process similar to the one reported in earlier papers \cite{Brachet92,Passot1995,agafontsev2015,agafontsev2017} (see Fig.\ref{Fig:Visu_early} (bottom row)).
A natural question arises, are those vortex sheets shrinking in a completely self-similar manner? If so, is such a self-similarity of the blowup (second kind) type as it is in the simplified turbulence models, e.g. the Leith model and the EDQNM model discussed in the introduction. This type of behaviour would correspond to a precursor of an eventual finite-time singularity of the inviscid dynamics. In the following, we will address these questions and study the self-similarity of the early stage evolution of our simulations.

A self-similar solution for the energy spectrum has the following form,
of the second kind (keeping notation of Nazarenko and Grebenev \cite{NazarenkoGrebenevLeith2017}):
\begin{equation}
E(k,t)= g(t)
F(\eta), \quad \eta=k/k_*(t), 
\label{Eq:SFSolution}
\end{equation}
where $k_*(t)$ is the characteristic wave number of the propagating front of the spectrum.  For the second-kind self-similarity, $g(t)=k_*(t)^{x}$ and $k_*(t)=(t_*-t)^b$, where  $x$ and $b$ are some similarity constants
which are to be found from solving a "nonlinear eigenvalue problem" of simultaneous matching the two boundary conditions: $F(\eta) \to \eta^{-x} $ for $\eta \to 0$ and $F(\eta) \to 0 $ (sufficiently fast) at $\eta \to \infty$. 
(Note that in such as solution, the spectrum is asympotically stationary at $\eta \to 0$).
Such kind of self-similarity was found 
for the Leith model \cite{connaughton2004warm,GNMCh2014,thalabard2015anomalous} and the EDQNM model \cite{BOS}, both being simplified models for 3D hydrodynamic turbulence.
This leads to a natural question if the second-kind kind self-similarity could also be observed in the DNS of the 3D Navier-Stokes turbulence.
As we will see later,  this form of self-similar solution is not observed in our simulations
which, however, does not rule out a possibility that it could appear in simulations at much higher resolution at
times very close to $t_*$. Our present results indicate
that the solution is also approximately self-similar with $ k_*(t)$  function of time increasing  exponentially. In this case, $t_*$ has a meaning of the time at which the self-similar stage ends (when the dissipative and smallest resolved scales are reached) rather than a "blowup" time. 

In order to study the pre-$t^*$ self-similar evolution and beyond, we fit the spectra with the following trial functions:

\begin{eqnarray}
E_{\rm fit\,  0}(k,t)&=& E_0(t) k^{-m_0(t)}e^{-2\delta_0(t) k}\label{Eq:Fit0}\\
E_{\rm fit\,  1}(k,t)&=& E_1(t) k^{-m_1(t)}e^{-2\delta_1(t) k-(\eta(t) k)^{n_\alpha}}\label{Eq:Fit1}
\end{eqnarray}
The interpretation is the following. At early times, the dissipation is negligible and in absence of finite time singularities, the velocity field is analytic and (at least) an exponential decay is expected in the high-$k$ range. In Euler flows, singularities appear as poles in the complex plane of the analytical continuation of the velocity flow \cite{SULEM1983138}. The scale $\delta(t)$ is usually called the analyticity strip and corresponds to the distance from the real plane to such singularities. If $\delta$ vanishes, then a singularity has crossed the real axes corresponding to an actual singularity of the flow.
At a later stage, when dissipation start acting, a stronger decay is expected. Then $\eta(t)$ would correspond to the Kolmogorov scale in fully developed turbulent states. We expect that at early times $\eta(t)$ will be very small, followed by a sudden increase at $t^*$. The proper value of $n_\alpha$ certainly depends on the kind of dissipation. It turns out that in our hyper-viscous setup the best fitting value is $n_\alpha=4$. 
{Note that self-similarity implies  $m_0=m_1=$~const, $\delta_0(t)=\delta_1(t) \propto 1/k_*(t)$. 
Also, as $\eta$ represents a dissipative scale, it can not decrease with time, and therefore self-simlarity imposes $\eta=0$, which is expected for $t<t^*$.} We will see that, like in the second-type self-similarity, our spectrum appears to be stationary at the IR end, which implies $E_0=E_1=$~const.
We will not fix the time dependence of $k_*$, and will find it from fitting the data. In this section we analyze the data of the run at resolution $2048^3$.

\begin{figure}[h!]
    \includegraphics[width=.9\textwidth]{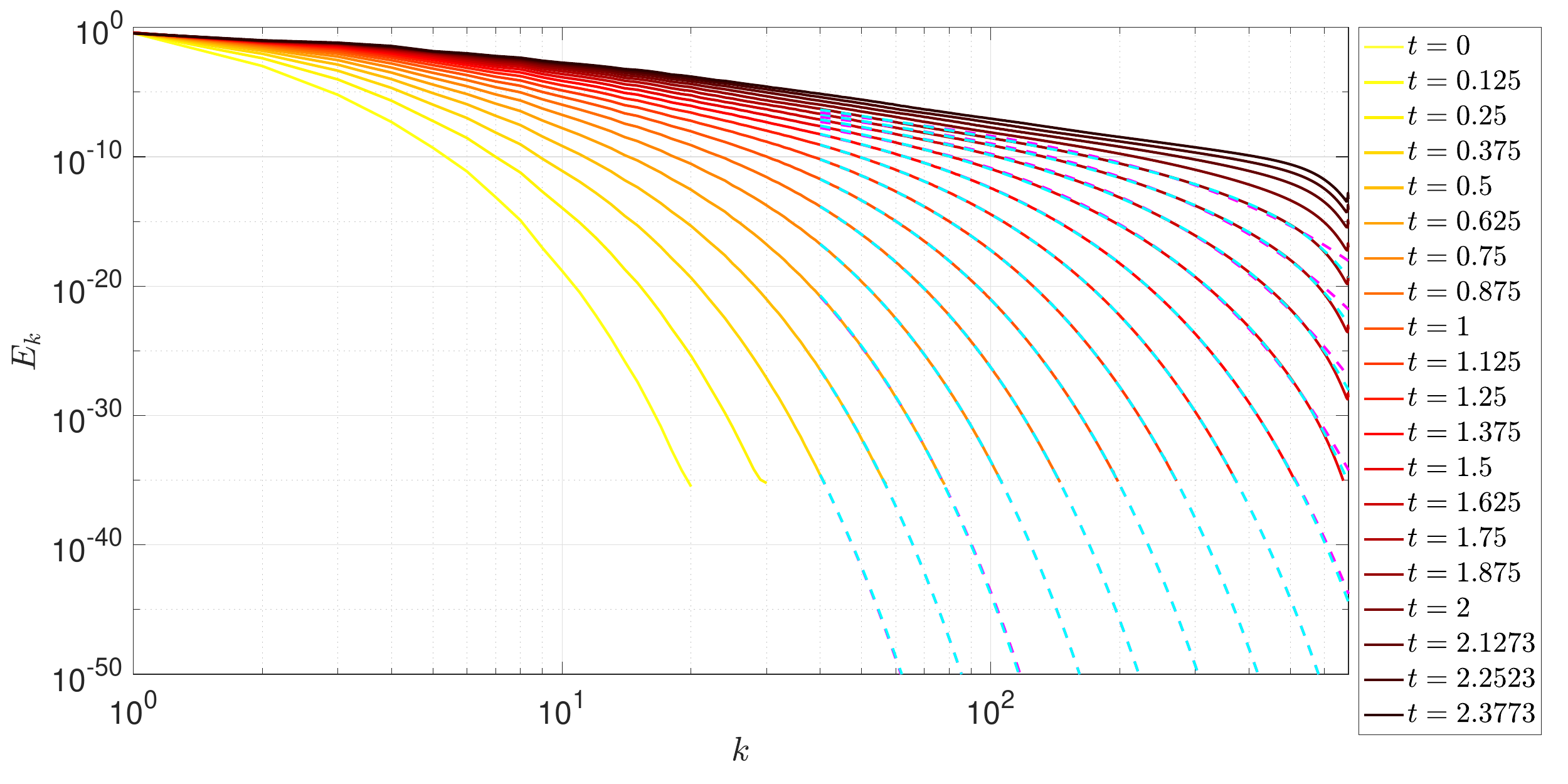}
    \caption{Temporal evolution of energy spectra and fits. Fit \ref{Eq:Fit0} is plotted in magenta and \ref{Eq:Fit1} in cyan. Run at resolution $2048^3$.}
    \label{Fig:FitresultsAndSpectra}
\end{figure}

\begin{figure}[h!]
    \includegraphics[width=.9\textwidth]{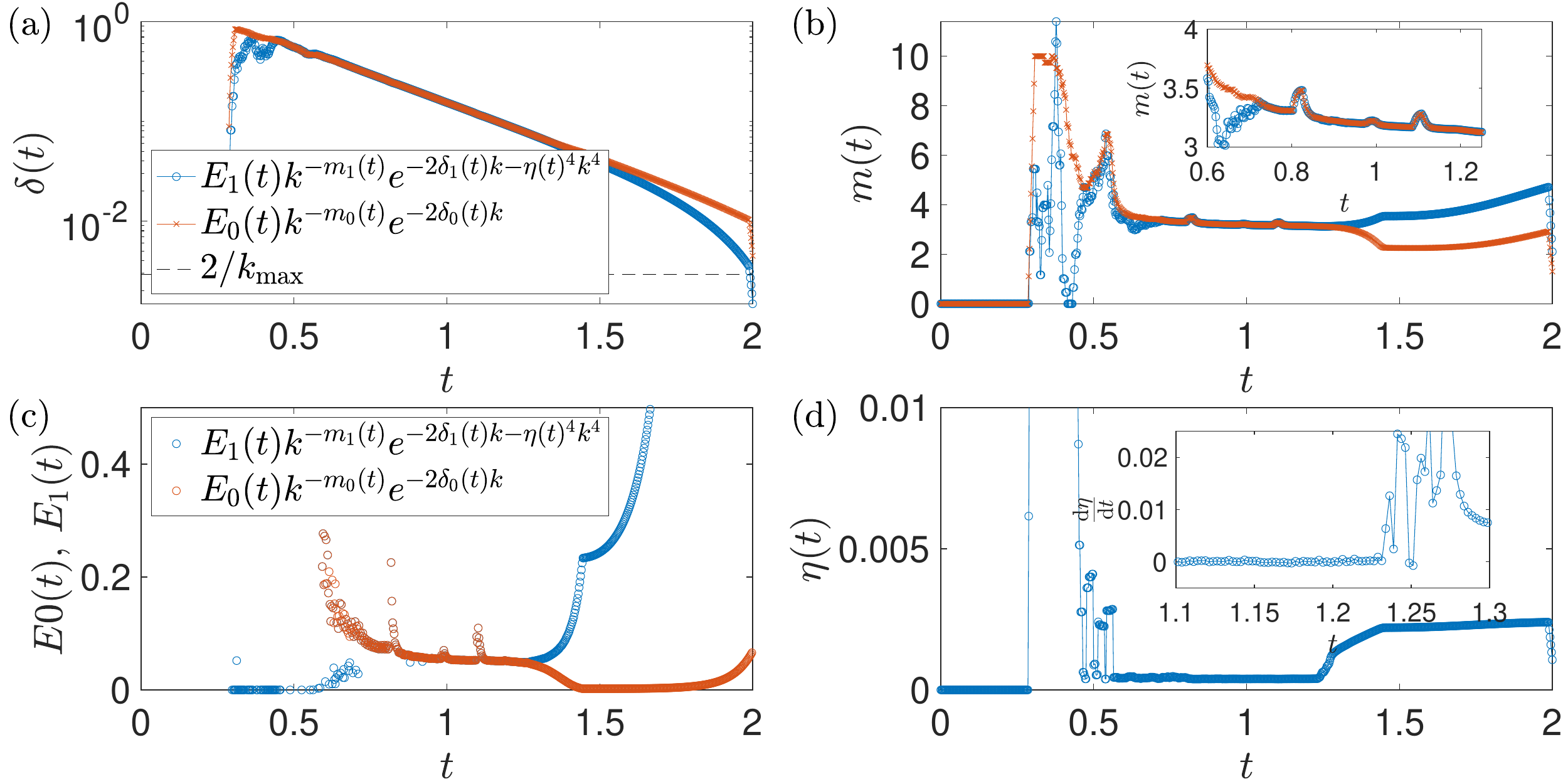}
\caption{Evolution of the fitting parameters in Eqs.\ref{Eq:Fit0} and \ref{Eq:Fit1}. Run at resolution $2048^3$.}
\label{Fig:Fitresults}
\end{figure}

The energy spectra at different moments of time, together with the fits \eqref{Eq:Fit0} and \eqref{Eq:Fit1} are shown in Fig.~\ref{Fig:FitresultsAndSpectra}.   
We see that, somehow, the fits could   be used  as an extrapolation of the numerical results beyond numerical precision. The temporal evolution of the  fitting parameters $E_{0,1}, \delta_{0,1}, m_{0,1} $ and $\eta$ are  displayed in Fig.~\ref{Fig:Fitresults}. 
From the inset of Fig.~\ref{Fig:Fitresults}.(d), which shows $\frac{{\rm d}\eta}{{\rm d}t}$,  we infer that the sudden jump of the derivative indicates that
$$
t_*\approx 1.24.
$$
Again, the suggested fits indicate approximate self-similarity only when the resulting parameters $E_0, E_1$ and $m_0, m_1$ are relatively constant in time.
From the evolution of the fitting parameters shown in Fig.~\ref{Fig:Fitresults}, we see that the period of time for which this condition is approximately satisfied is $0.6 \lesssim t \lesssim t_*$.
From the lower-left frame of Fig.~\ref{Fig:Fitresults}, we see that at the pre-$t_*$ stage the exponent of the power law behind the propagating front of the spectrum is $m \sim 3.25$, which is consistent with the previous numerical results obtained at low resolutions in  \cite{Brachet92}.

Let us seek self-similar solutions by determining the temporal evolution of $k_*(t)$ numerically.
We can define $k_*(t)$ by using the fits Eqs.~\ref{Eq:Fit0} and \ref{Eq:Fit1}  or by  measuring it directly from the spectrum, namely
\begin{eqnarray}
k_*^{\rm raw}(t)&=&\max{[k\,| E(k,t)> \epsilon}]\\
\hbox{or} \quad k_*^\delta(t)&=& \lambda/\delta(t).
\end{eqnarray}
We expect that, up to a constant, all definitions present the same trend. In numerics we have set $\epsilon=10^{-30}$. The pre-factor $\lambda$ is set in order to match the values of $k_*^{\rm raw}$ and $k_*^\delta(t)$ during the self-similar evolution.

\begin{figure}[h]
\includegraphics[width=.48\textwidth]{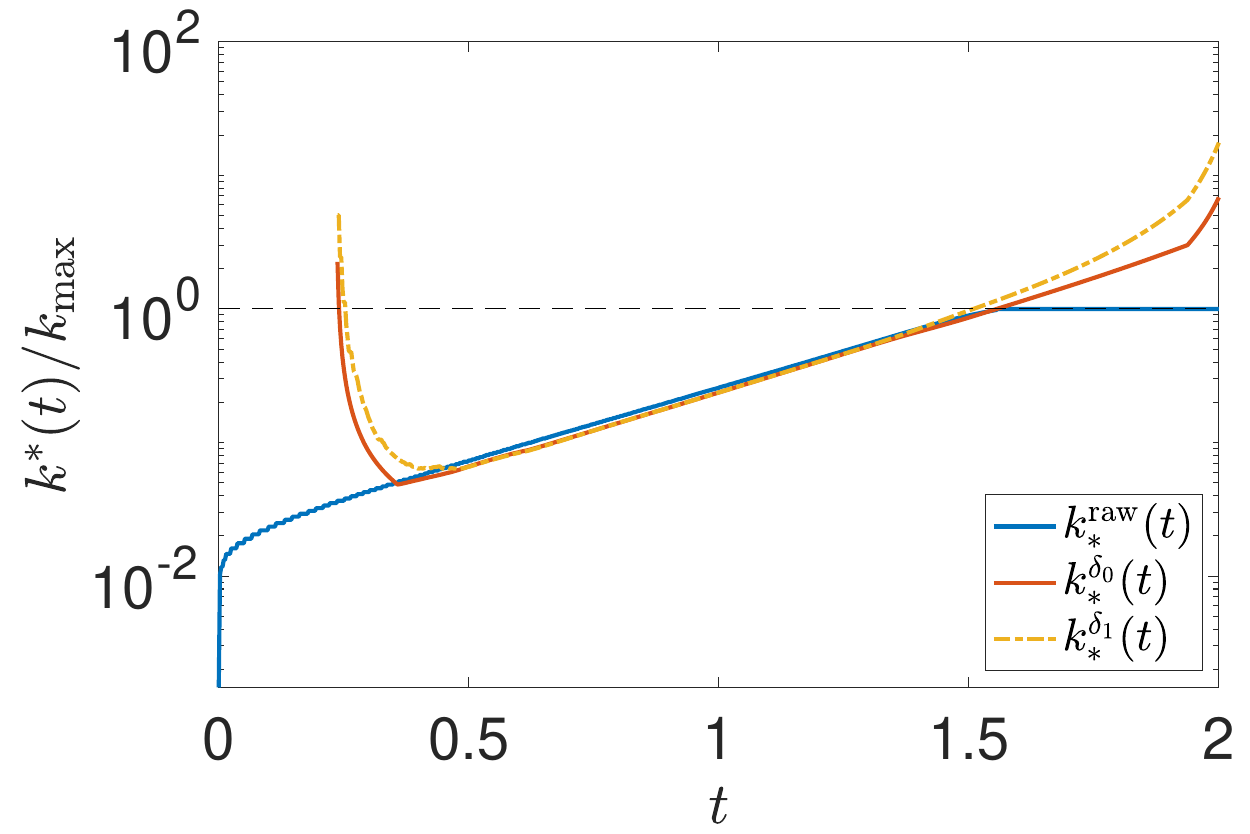}
\caption{Temporal evolution of $k^*t$. The maximum wavenumber of the simulation is $k_{\rm max}=2048/3$.}
\label{Fig:Kstar}
\end{figure}

In Fig.~\ref{Fig:Kstar} we present the temporal evolution of  $k_*(t)$, normalized by the maximum wavenumber of the simulation $k_{\rm max}$.
We observe a clear exponential growth of $k^*$ in the range $k^*(t)<k_{\rm max}$, which, at the current resolution, excludes the second-kind self-similarity with a finite-time singularity scenario. 
Again, we do not rule out a theoretical possibility that at much higher resolution one might see a transition to an accelerated (second-kind) self-similar regime very close to  $t_*$.

On the other hand, the exponential growth of $k^*(t)$, and the respective exponential shrinking of characteristic physical scale, has a clear physical interpretation as a process of shrinking of the pancake-like vorticity structures embedded to a strong strain field produced by the large scales.
The fact that the large-scale strain is indeed very strong is explained by a very steep initial energy spectrum. For  a very strong strain, the action of the pancakes onto themselves can be neglected compared to the action of the large-scale strain on them. Such a nonlocal (in scale) dynamics can be described the Rapid Distortion Theory (RDT) \cite{rdt,nazarenko_kevlahan_dubrulle_1999}, and the exponential shrinking of the scale in this approach is simply due to the motion of the vorticity along the exponentially converging steamlines of the large-scale straining motion. 

\begin{figure}[h!]
\includegraphics[width=.8\textwidth]{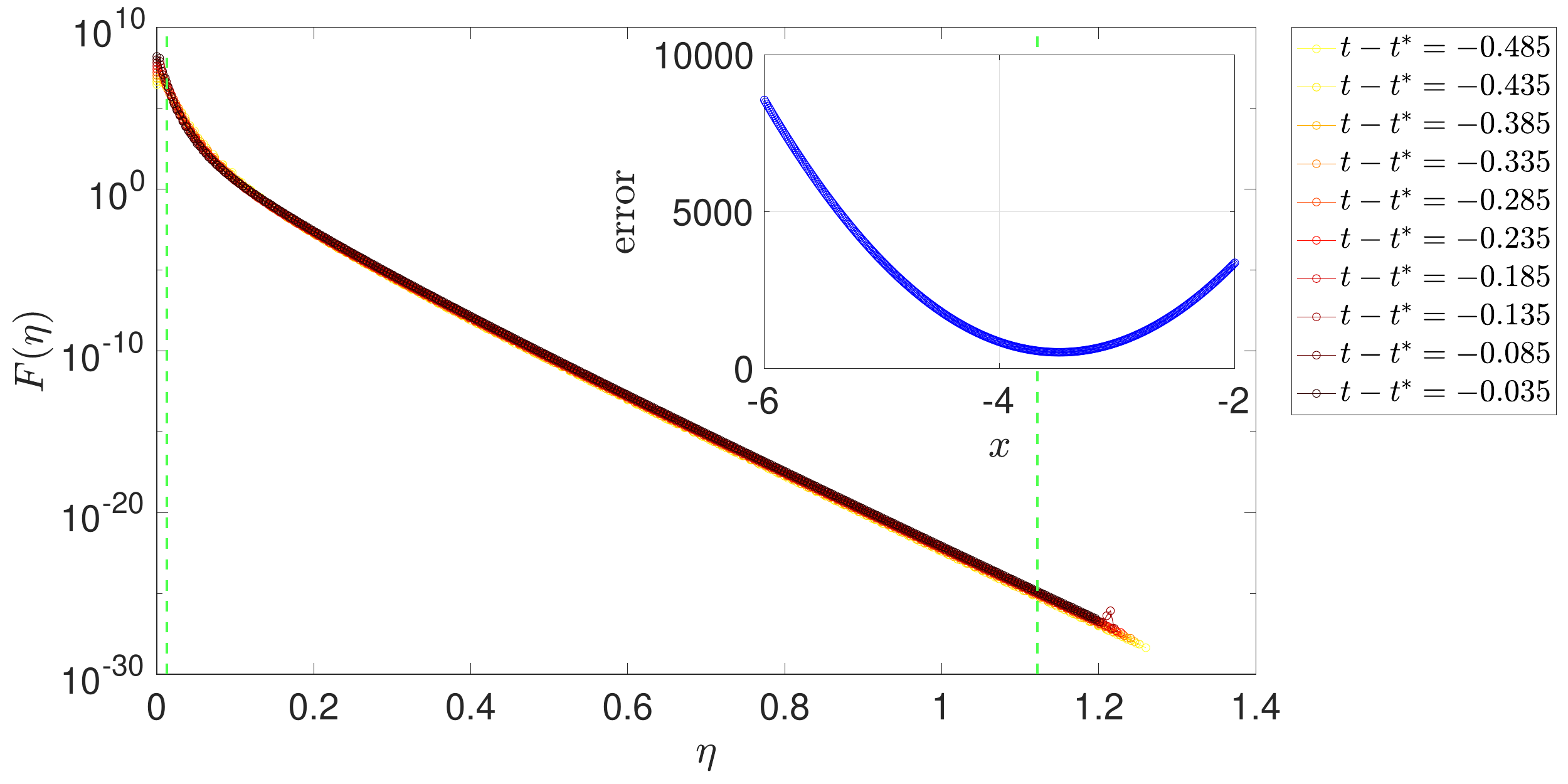}
\caption{ Self-similar form Eq.~\eqref{Eq:SFSolution} for different times before $t *$. Error function (arbitrary normalization) \eqref{Eq:error} as a function of the scaling parameter $\gamma$. Bounds of the integrals are set to $\eta_{\rm i}=0.0126$, $\eta_{\rm f}=1.12$ (green dashed vertical lines) and  $T_0=.7475$ , $T_1=1.2275$. Run at resolution $2048^3$.}
\label{Fig:sefsilmarspectra}
\end{figure}

Let us now embark on a more formal search of the self-similar solution that provides the best fit to our numerical data before $t_*$.
We look for a self-similar solution given by equation \eqref{Eq:SFSolution} with $g(t)=k_*(t)^x$.
We use  $k_*(t)$ found from fitting the data for finding the scaling parameter $x$ via minimizing the following error,
 \begin{eqnarray}
error(x)&=&\int_{T_0}^{T_1}\int_{T_0}^{T_1}\left[ \int_{\eta_{\rm i}}^{\eta_{\rm f}} \left(     \log{\tilde{F}(\eta,t)} - \log{\tilde{F}(\eta,s)}    \right)^2{\rm d}\eta\right]{\rm d}s{\rm d}t\label{Eq:error}\\
\eta=\frac{k}{k_*(t)}&,&
\tilde{F}(\eta,t)=k_*(t)^{-x} E(k,t).
\end{eqnarray}
Note that for a perfect self-similar evolution,  $\tilde{F}(\eta,t)=F(\eta)$, and then for the correct value of the exponent $x$ we thus have $error(x)=0$.
The resulting error of minimizing \eqref{Eq:error} is displayed on the inset of figure \ref{Fig:sefsilmarspectra}, and gives the optimal value $x=-3.49$. The error is computed in the interval $\eta\in(0.0126,1.1220)$ (represented by dashed green vertical lines). The value of the exponent $-3.49$ is consistent withithin $7\%$ with the result for the power-law exponent behind the propagating front $-m \approx 3.25$, as seen in the inset of Fig.~\ref{Fig:Fitresults} (b).
We use now the obtained value of $x$ to show the self-similar behavior of the energy spectra. The function $F(\eta)$ at different times is displayed in Figure \ref{Fig:sefsilmarspectra}. The collapse  onto a single curve is excellent.

\section{Intermediate times, rolling pancakes and vortex ribs}
\label{sec:int}

\begin{figure}[h!]
            \includegraphics[width=.99\textwidth]{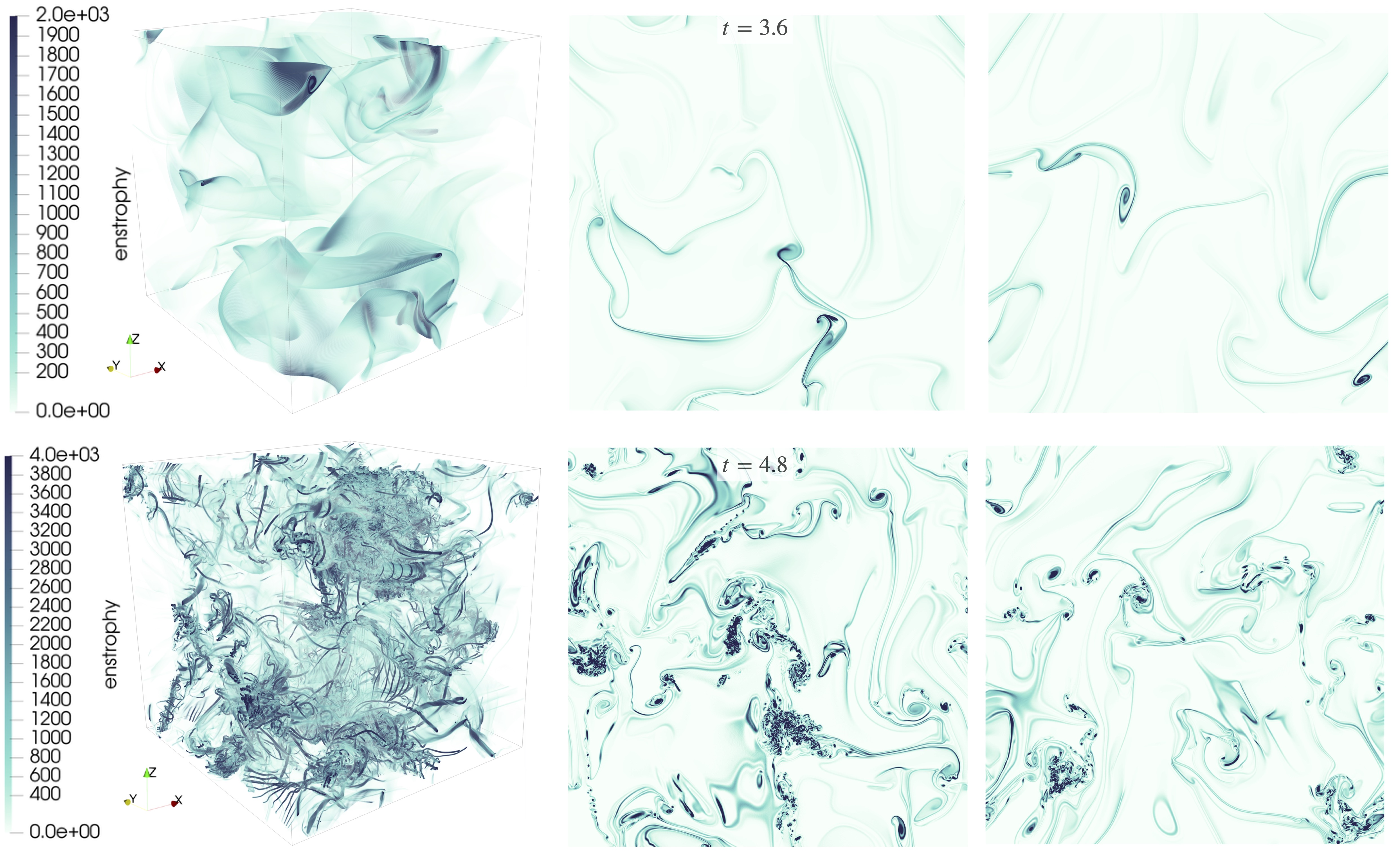}
            \includegraphics[width=.99\textwidth]{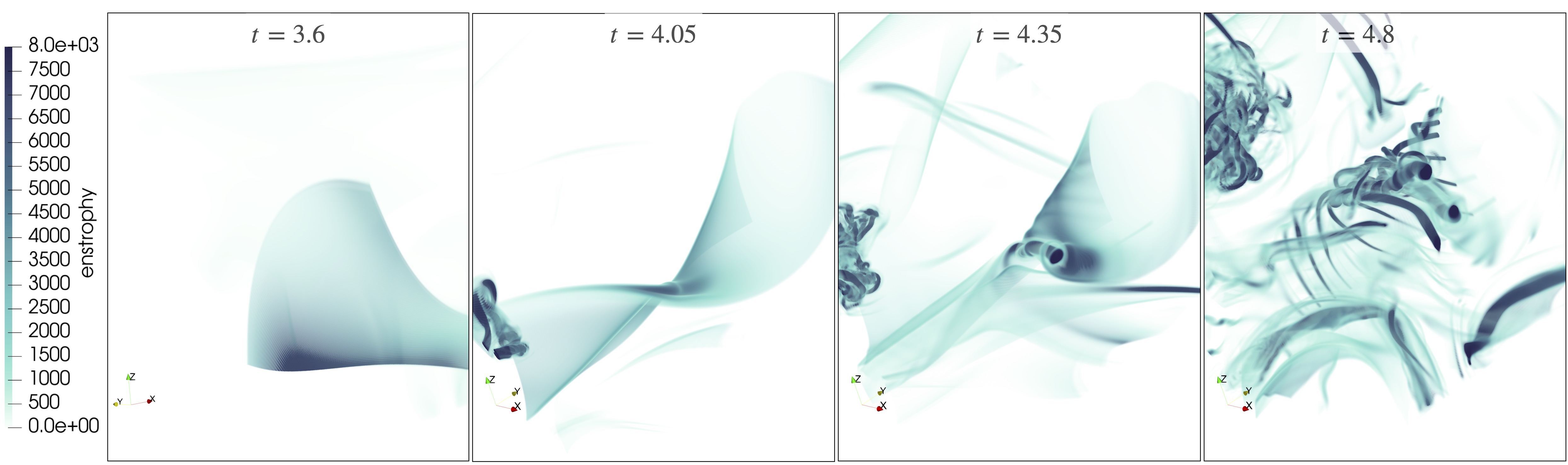}
            \caption{Visualizations of the enstrophy field at $t=3.6$ (top row) and $t=4.8$ (middle row). The left panels of the top and the middle  rows show 3D visualizations of the field, whereas the center and right columns display cuts of the $xy$- and $xz$-planes.
            The bottom row show several consecutive frames zoomed into a region with a typical vortex sheet roll-up and formation of the vortex ribs.  Run at resolution $1024^3$.}
            \label{Fig:Vis_ribs}
        \end{figure}

After $t^*$, dissipation start acting, taking the system  beyond the purely self-similar shrinking of pancakes. Visualizations of the enstrophy field at $t=3.6$ and $t=4.8$ are displayed in Fig. \ref{Fig:Vis_ribs}. 
Early post-$t_*$ evolution is characterised by persistence of the vortex pancakes which continue to shrink and whose vorticity continues to grow for a time of about three to four $t_*$. In the top row of Fig. \ref{Fig:Vis_ribs} we see that at
$t=3.6$ some of the vortex pancakes are rolling up at the edges, 
 quickly exciting small scales which are purged by hyperviscosity. Such a purging allows the pancakes  to keep shrinking in other (more central) areas 
 until the hyper-viscosity counteracts with the shrinking by (hyper-) diffusing it out of the thin layers thereby leading to a slowdown and eventual arrest of the shrinking process.  
  After that some pancakes become unstable which leads to their breakup into sets of quasi-periodic quasi-colinear thin vortex tubes, ``vortex ribs", see the middle row of Fig. \ref{Fig:Vis_ribs} for the enstrophy field at
$t=4.8$. 
Transverse cross-section of such ribs shows a vorticity pattern typical for a vortex street arising after the Kelvin-Helmholtz type instability
of a vortex layer. Similar behaviour was previously observed in $256^3$ simulations of Navier-Stokes equations, and the rollup of the vortex pancakes into the array of vortex filaments was explained as a result of instability which could be understood as an instability of the Burgers vortex layer \cite{Passot1995}. The dissipation mechanism in our simulations is hyperviscosity rather than a regular viscosity, but clearly similar quasi-stationary vortex layers in which the vortex stretching by external large-scale strain is balanced by a layer (hyper-)diffusion must appear and become unstable in our system too. 
After forming, vortex ribs start approaching to each other and form  twisted vortex bundles reminiscent of ropes, as also seen in the middle row of Fig. \ref{Fig:Vis_ribs} (left panel). 
The bottom panel of Fig.~\ref{Fig:Vis_ribs} shows a zoom of the box where the shrinking of pancakes, the rolling and the formation of vortex ribs is clearly visible.
Further in time, the number of vortex ribs and ropes proliferate and become abundant throughout the fluid, and their mutual interactions lead to an increasingly chaotic vortex tangle at $t \gtrsim 4.5 t_*$, see e.g. the transition in the enstrophy field character from $t=4.8$, the bottom row of Fig. \ref{Fig:Vis_ribs},   to $t=7.05$ in Fig. \ref{Fig:visuLaterTimes}.

In terms of the energy spectrum, the evolution for $t_* < t <4.5 t_*$ is not too eventful: it is characterised by a gradual increase of the power-law exponent from $\sim -3.4$ to $\sim -5/3$.
\begin{figure}[h!]
    \includegraphics[width=.9\textwidth]{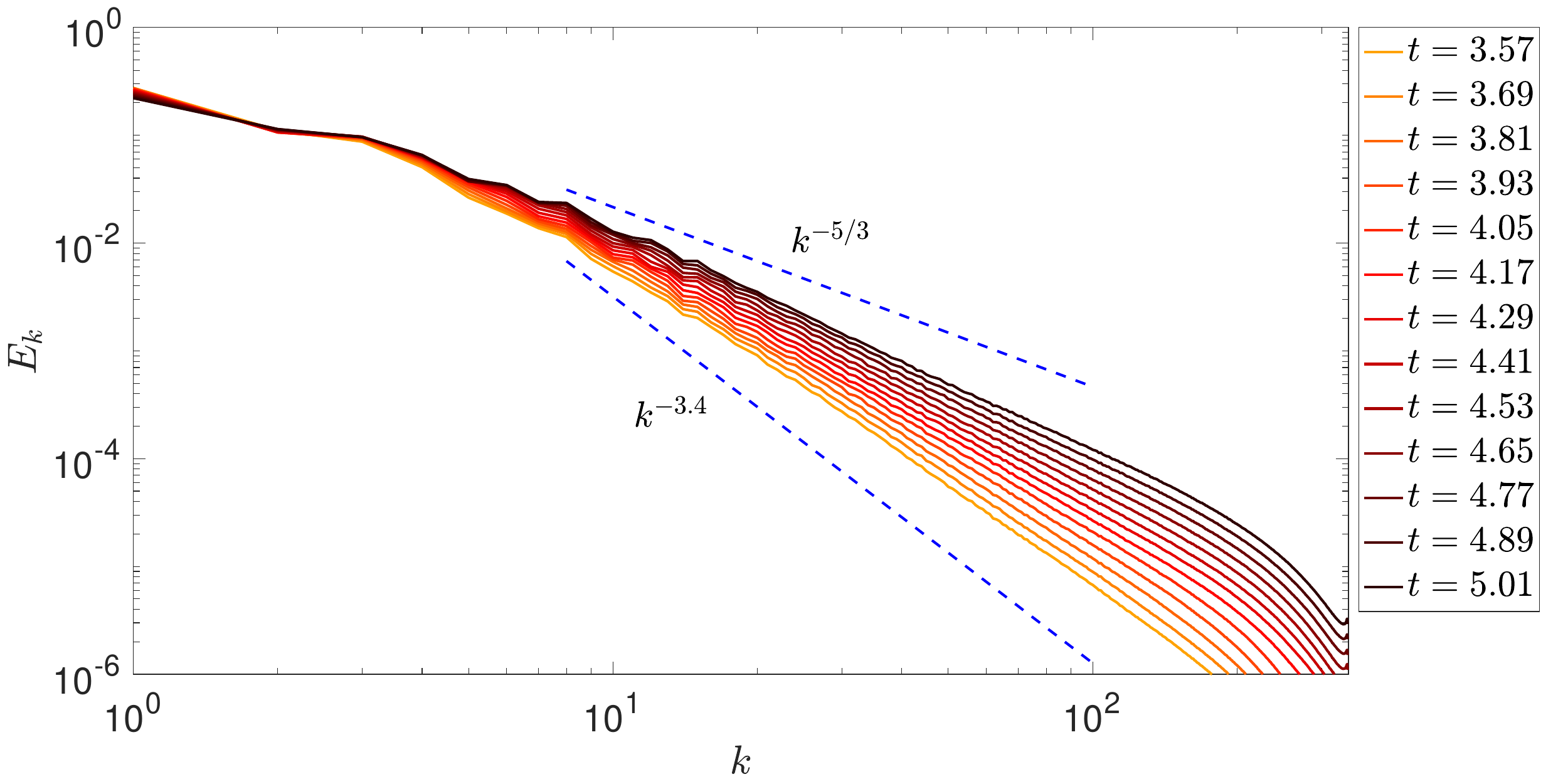}
    \caption{Energy spectra at intermediate times. Run at resolution $1024^3$}
    \label{Fig:SpectraInterTimes}
\end{figure}

\section{Long times}
\label{sec:last}

\begin{figure}[h!]
    \includegraphics[width=.99\textwidth]{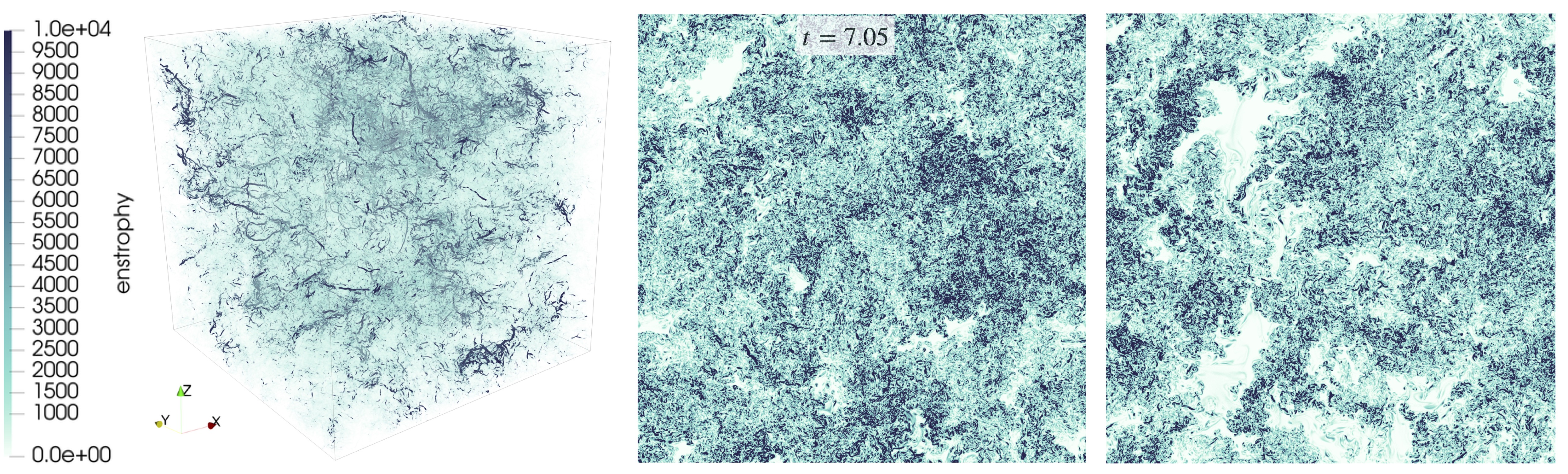}
    \caption{Visualizations of the enstrophy field at $t=7.05$. The left panel shows a 3D visualization of the field, whereas the center and right columns display cuts of the $xy$- and $xz$-planes. Run at resolution $1024^3$.}
    \label{Fig:visuLaterTimes}
\end{figure}

\begin{figure}[h!]
    \includegraphics[width=.99\textwidth]{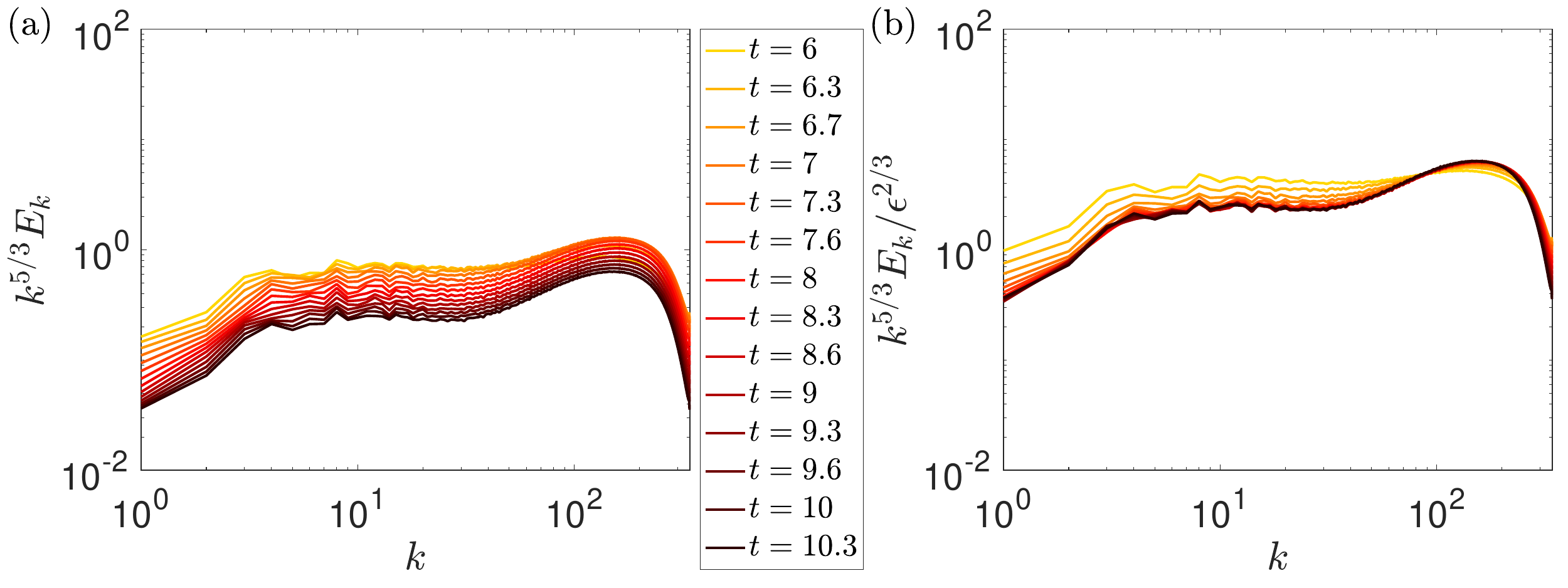}
    \caption{Temporal evolution of compensated energy spectrum at larger times. Run at resolution $1024^3$.}
    \label{Fig:K41}
    \end{figure}

We now address the later stage, $t \gtrsim 7$, which corresponds to the dynamics after the maximum of dissipation is reached (see Fig.~ \ref{Fig:EnerAndDisp}). At this stage, vortex ribs are not longer present and filaments and vortex structures display a more chaotic behviour as apparent in Fig.~\ref{Fig:visuLaterTimes}, tipical of turbulent flows.
Indeed, turbulence is fully developped and the energy spectrum displaus a Kolmogorov $-5/3$ scaling as shwon in Fig.~\ref{Fig:K41}. 
We notice the emergence of a clear Kolmogorov scaling for $5<k<50$. This is apparent in the left panel of Fig.~\ref{Fig:K41} where $k^{5/3}$-compensated spectra are displayed. For $k>50$ we see a bump which is typical in simulations with hyperviscosity: it corresponds to a tendency to the turbulence thermalization due to presence of the cutoff wave number $k_{max}$. The right panel of Fig.~\ref{Fig:K41} displays the spectra compensated by $\varepsilon(t)^{2/3}k^{5/3}$. Here we can see that the whole spectrum, including both the Kolmogorov and the thermal bump parts, follow approximately the same law of decay at the late stages. In the other words, the late stage of the spectrum evolution is also self-similar. 

\begin{figure}[h!]
    \includegraphics[width=.475\textwidth]{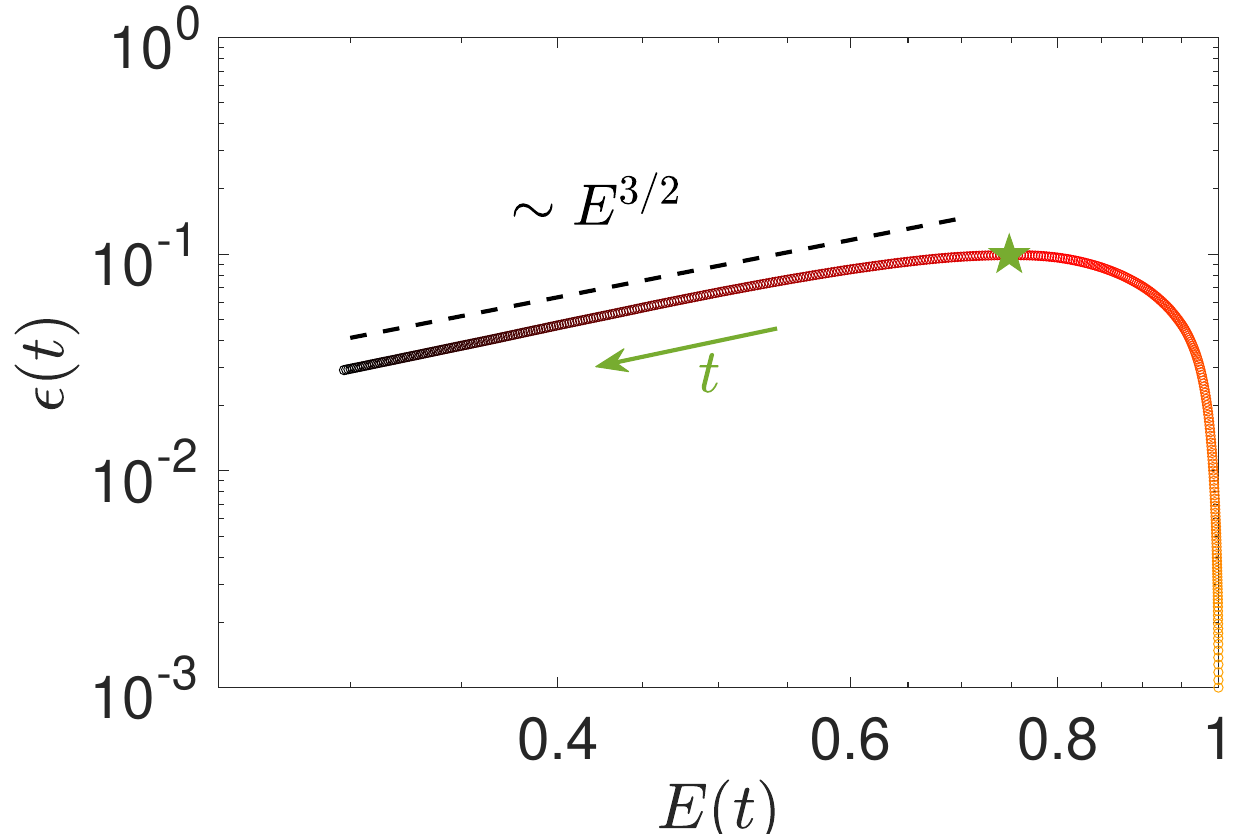}
    \caption{Temporal evolution of compensated energy spectrum at larger times. Run at resolution $1024^3$.}
    \label{Fig:DecayLaw}
\end{figure}

 {The decay of energy and dissipation observed in Fig.~\ref{Fig:EnerAndDisp} is typical of turbulent flows. Several power laws have been proposed and reported depending on the nature of the large-scale configuration of the turbulent flow \cite{Lin,Saffman,NazarenkoGrebenevLeith2017,panickacheril2022laws}. Turbulent decay laws are constructed under the assumption of  existence of an inertial range. Assuming that the energy dissipation is equal to the energy input at large scales, a simple balance equation can be easily derived $\epsilon=-\frac{dE}{dt}\sim E^{3/2}/\ell$, where $\ell$ is the integral scale of the flow. In general, the difficulty lies in determining the temporal evolution of the integral scale $\ell$. Such phenomenological arguments also apply to our hyperviscous simulation as they only require a well-developed turbulent flow obeying Kolmogorov phenomenology. In our particular case, the situation is simpler as the integral scale is saturated at the size of the box, and therefore $\ell$ remains constant in this stage. This assumption leads to a decay low $E(t)\sim t^{-2}$. In our simulations, the decaying range is too narrow to check this exponent directly, but our data is compatible with the law $\epsilon(t)\sim E(t)^{3/2}$, as shown in Fig.~\ref{Fig:DecayLaw}.}

\section{conclusions}

In this paper, we have performed DNS of the hyper-viscous Navier-Stokes system with no forcing and with only few randomly selected low-wavenumber Fourier modes excited initially. The goal was to study transition to the fully turbulent state, with specific attention to self-similar behaviour and to different coherent quasi-singular vortex structures appearing at different evolution stages. We we motivated by the previous results for the differential (Leith) and integro-differential closures (EDQNM) for the hydrodynamic turbulence as well as for the Burgers equation. For these systems, the second-kind self-similarity was observed, where the propagating spectral front reached the infinite wavenumber in a finite time $t^*$ after which a reflected wave propagates toward the smaller wavenumbers bringing the Kolmogorov scaling in the wake (Burgers scaling in the Burgers equation model). In the Burgers case this evolution is related to the successive appearance of singularities (quasi-singularities in presence of small viscosity) of two types: the pre-shock and the shock respectively. Thus we wanted to know if there is a self-similar evolution in the Navier-Stokes system, if this self-similarity is of the second (blowup) type (with a consequence of a blowup occuring in the Euler equations), and if we could identify quasi-singular structures of different types during the evolution en route to the Kolmogorov state.

Our analysis of the DNS data shows that at the inviscid stage the evolution is indeed self-similar but it not of the second kind: instead of the blowup propagation of the spectral front to infinite $k$ in a finite time we observe an exponential in time motion. Second, we do observe quasi-singularities: (1) shrinking vortex pancakes are forming at the inviscid stage via an RDT-type mechanism of vortex stretching/shrinking by a large-scale strain and (2) breakup of pancakes into quasi-periodic sets of quasi-singular vortex filaments, vortex ribs, by a Burgers vortex sheath instability. The last stage of transition to the Kolmogorov state is characterised by ribs twisting into ropes followed by entanglement of the ribs and ropes into a visually unstructured vortex tangle.

There is a natural question: why do the turbulence closures, Leith model and EDQNM, both exhibit the second-kind self-similarity with a blowup spectral front propagation whereas our DNS of the hyper-viscous Navier-Stokes does not? The clue can be found in the RDT-type mechanism of the vortex pancake stretching/squashing. Indeed, this mechanism is based on a nonlocal (in the scale space) interaction: the strong large-scale strain affects the small vorticity scales associated with the thin pancakes much stronger than the small scales affect each other. But the Leith  and EDQNM models do not capture such effects: locality of interactions is "built into" these two models. Thus we arrive at a conjecture that it is the nonlocal interactions of scales that upset the blowup behaviour. However, at this point it could not be ruled out that much higher resolution simulations, or simulations with some special initial conditions, might reveil a blowup  which could supersede the exponential scale shrinking.

\section{Acknowledgements}
This work was funded by the Simons Foundation Collaboration grant Wave Turbulence (Award ID 651471). Computations were  carried out at the M\'esocentre SIGAMM hosted at the Observatoire de la Co\^ote d'Azur.


%

\end{document}